\documentclass[onecolumn,showpacs,preprintnumbers,amsmath,amssymb,nofootinbib]{revtex4}

\usepackage{graphicx}
\usepackage{dcolumn}
\usepackage{bm}
\usepackage{graphics}
\usepackage{epsfig}
\usepackage{color}
\usepackage{amssymb}
\usepackage{amsmath}
\usepackage{amsfonts}

\begin{document}
\title{Finite Pressure Corrections the to Parton Structure of Baryon Inside a Nuclear Medium.}

\author{Jacek Ro\.zynek} \email{rozynek@fuw.edu.pl}
\affiliation{The So\l tan Institute for Nuclear Studies, Ho\.za
69, 00-681 Warsaw, Poland}




\begin{abstract}
 Our model calculations performed in
the frame of the Relativistic Mean Field (RMF) approach show how
important are the modifications of baryon Structure Function (SF)
in Nuclear Matter (NM) above the saturation point. They originated
from the conservation of a parton longitudinal momenta - essential
in the explanation of the EMC effect at the saturation point of
NM. For higher density the finite pressure corrections emerge from
the Hugenholtz -van Hove theorem valid for NM. The density
evolution of the nuclear SF seems to be stronger for higher
densities. Here we show that the course of Equation o State (EoS)
in our modified Walecka model is very close to that obtained from
extensive DBHF calculations with a Bonn A potential. The nuclear
compressibility decreases. Our model - a nonlinear extension of
nuclear RMF, has no additional parameters. Modelling deep
inelastic scattering on nuclear, neutron or (strange) matter with
finite pressure, we attempt to predict also the change of baryon
masses in a strange nuclear medium. These changes are derived from
the Momentum Sum Rule (MSR) of quark longitudinal momenta for
different constituents. The increasing pressure between baryons
starts to increase baryon Fermi energies $e_F$ in comparison to
average baryon energies $e_A$ , and consequently the MSR is broken
by the factor $e_F/e_A$ from the hadron level in the convolution
model. To compensate this factor which increases the longitudinal
momentum for nuclear partons, the baryon SF in the nuclear medium
and their masses have to be adjusted. Here we assume that,
independently from density, quarks and gluons carry the same
amount of longitudinal momenta - the similar assumption is used in
the most nuclear models with parton degrees of freedom.
\end{abstract}
 \maketitle
\section{The nuclear deeply
inelastic limit - nuclear equilibrium}

In the nuclear deep inelastic scattering on nuclei our
time-distance resolution is given by variable z\cite{Jaffe}:
\begin{equation}
z = 1/(xM_N)
\end{equation}
which measures the propagation time of the hit quark caring the
$x$ fraction of the longitudinal momentum of the nucleon of mass
$M_B$. Start with the scenario where the partonic mean free paths
$z$ are much shorter then the average distances between nucleons.
This means that partons (inside mesons) remain inside the "volume"
of a given nucleon and therefore we can treat nucleons as
noninteracting objects remaining on the energy shell not affected
by neighboring nucleons.

In the light cone formulation\cite{Jaffe,Fran}, $x_A$ corresponds
to the nuclear fraction of quark longitudinal momentum
$p^+_q=p_q^0+p_q^3$ and is equal (in the nuclear rest frame) to
the ratio $x_A=p_q^+/M_A$. But the composite nucleus is made of
hadrons which are distributed with longitudinal momenta $p_h^+$,
where $h=N,\pi,...$ stands for nucleons, virtual pions, ... . In
the convolution model\cite{Jaffe,Fran} a fraction of parton
longitudinal momenta $x_A$ in the nucleus is given as the product
$x_A=x_h*y_h$ of fractions: parton momenta in hadrons $x_h\equiv
Q^2/(2M_h\nu)= p_q^+/p_h^+$ and longitudinal momenta of hadrons in
the nucleus $y_h=p_h^+/M_A$. The nuclear dynamics of given hadrons
in the nucleus is described by the distribution function
$\rho^h(y\equiv y_h)$ and SF $F^h_2(x\equiv x_h)$ describes its
parton structure. In the convolution model restricted to nucleons
and pions (lightest virtual mesons) the nuclear SF $F^{A}_{2}$ is
described by the formula:
\begin{equation}
F^{A}_{2}(x_A)\!=\!\!\!\int\!\!{ydy}\!\!\!
\int\!{\!dx}\delta(x_A\!-\!xy)
(\rho^N\!(y)F^{B}_{2}\!(x)+\rho^{\pi}\!(y)F^{\pi}_{2}(x))
\label{structure}
\end{equation}
where $F_2^{\pi}$ and $F_2^{B}$ are the parton distributions in
the virtual pion and in the bound nucleon. The nucleon
distribution $\rho ^{A}$ in the basic convolution formula can be
simplified in the RMF to the form\cite{Mike}:
\begin{eqnarray}
\rho^{A}(y)\!&\!=\!&\!\frac{4}{\rho}\int_{|p|>p_F}\frac{S_N(p)d^3p}{(2\pi)^3}
(1\!+\!\frac{p_3}{E^*_p})\,\delta(y\!-\!p^+/\varepsilon_N)
\nonumber
\\&=& \frac{3}{4}\left(\frac{\varepsilon_N}{k_F}\right)^3 \left[
\left(\frac{p_F}{\varepsilon_N} \right)^2-\left(y-\frac{E_F}{
\varepsilon_N} \right)^2 \right], \label{RMF}
\end{eqnarray}
Here the nucleon spectral function was taken in the impulse
approximation: $S_N=n(p)\delta ( p^o-(E^{^{*}}(p) +U_V))$ and
$E^{^{*}}(p)=\sqrt{M_N^2+p^2}$. $E_F$ is the nucleon Fermi energy
and $y$ takes the values given by the inequality
$(E_F-p_{F})/\varepsilon_N<y<(E_F+p_{F})/\varepsilon_N$.

The MSR   for the nucleonic part is sensitive to the Fermi energy
as can be seen from the integral:
\begin{equation}
\int\!dy\,y\rho^{A}(y)=\frac{E_F}{\varepsilon_N} \label{RMF2}
\end{equation}
Thus the nucleonic part of MSR  gives a factor $E_F/\varepsilon_N$
which is equal to $1$ at the saturation point. If the nucleon SF
is not changed in the medium (no EMC effect except Fermi motion)
the total (\ref{SR}) MSR  is satisfied without nuclear pions:
\begin{eqnarray}
\frac{1}{A}\! \int\!\!F^{A}_{2}(x)\!dx\!=\!\!\int\!dyy\rho
^{A}(y)\!\int\!\! F^{N}_{2}\!\left(x\right)\!dx\! \! =\!\!\int\!\!
F^{N}_{2}\!\left(x\right)\!dx  \label{MSR}
\end{eqnarray}

 Summarizing, good description \cite{jacek,RW} of these deeply inelastic processes
 without gluon degrees of freedom allows us to
assume that fraction of momentum carried by quarks
 does not change from nucleon to
nucleus ($\sim$ one half, the rest is carried by
 gluons).  We will assume that balance also above the saturation point of NM.
 Now for the non zero pressure the Fermi energy in NM
 is no longer equal to the average binding energy, and corrections to MSR (\ref{RMF2})
 proportional to the pressure which will now be investigated.

\section{Non-equilibrium correction to nuclear distribution.}

For finite pressure very important is well known Hugenholtz van
Hove relation between $E_F$, $\varepsilon_N$ and pressure $p$ (see
for example) \cite{kumar}. The Fermi energy is defined as density
derivative of the total nuclear energy $E=A\varepsilon_N$:
\begin{eqnarray}
E_F=\frac{d}{d\varrho}\left(\frac{E}{\Omega} \right) \nonumber \\\
E_F= \varepsilon_N+\varrho\frac{d\varepsilon_N}{d\varrho}
\end{eqnarray}
where $A/\varrho=\Omega$ gives the volume. At the saturation point
$E_F = \varepsilon _A$. But for negative pressure $p$
\begin{equation}
\int\!dy\,y\rho^{A}(y)=\frac{E_F}{\varepsilon_N}<1. \label{RMF3}
\end{equation}
and we have room for additional pion inside NM (even with
unmodified nucleons) in the Bj\"orken limit (parton picture).

\subsection{Positive pressure.}
Consider the additional pion contributions. The amount of $1\%$ of
the total nuclear momentum\cite{jacek,RW,drell,ms2} was estimated
 from (\ref{sum}) for $x>x_L$ due to the smaller, $x$
dependent nucleon mass $M_B(x)$. For higher density, the average
distances between nucleons are smaller, therefore parameter
$x_{L}$ (\ref{eq:mamam2}) will increase with density. It means
that
 the room for nuclear pions given by
(\ref{sum}) from $x$ dependent nucleon mass $M_{B}$  will be
reduced for higher densities and a dependence of $M_B$ from
Bj\"orken $x$ will vanish gradually. Also the pion effective cross
section is strongly reduced at high nuclear densities above the
threshold in $N+N=N+N+\pi$ reaction calculated in Dirac-Brueckner
approach \cite{hm1} ( also with RPA insertions to self energy of
$N$ and $\Delta$ \cite{oset} included).
 Therefore for positive pressure the nuclear pions carry
 much less then $1\%$ of the nuclear longitudinal momentum and dealing
with a non-equilibrium correction to the nuclear distribution
(\ref{structure}) we will restrict considerations to the nucleon
part without additional virtual pions between them
\begin{eqnarray}
\frac{1}{A}F^{A}_{2}(x) dx = \frac{1}{A}\int dy \rho ^{A}(y) \int
F^{N}_{2}\left(x/y\right).  \label{nucleon}
\end{eqnarray}
  The Equation of State (EOS) for NM has to match
the saturation point with compressibility
$K^{-1}=9\varrho^2\frac{d^2}{d\varrho^2} \frac{E}{A}$ but then the
behavior for higher densities is different for different RMF
models. We compare here two extreme examples: stiff model of
Walecka \cite{wa} and  nonrelativistic expansion in powers of
Fermi momentum \cite{dabrnuc}. For linear coupling in the standard
Walecka model at saturation density of NM compressibility is too
large ($K^{-1}\simeq560$ MeV). The energy $E_{press}=p/\varrho$
shown in Fig.\ref{meson} influences the nuclear EOS. In both
models, two coupling constants of the theory are fixed by the
empirical saturation density. Our approach is different. In this
work  we consider the change of the nucleon mass with the change
of the parton distribution (nucleon SF) above the saturation
point. The increasing pressure between nucleons starts to increase
the $E_F$ (\ref{RMF2}) and consequently the sum rule (\ref{MSR})
is broken by the factor $E_F/\varepsilon_N>1$. To compensate this
factor which increases the longitudinal momentum (\ref{MSR}) of
nuclear partons, the nucleon SF in the nuclear medium has to be
changed. For good estimate, in order to proceed without new
parameters, assume (similarly to (\ref{scale})) that the changes
of SF will be included through the changes of Bj\"orken $x$ in the
medium. Multiplying the argument of the SF by a  factor
$E_F/\varepsilon_N$ the SF will be squeezed towards smaller $x$
and the total fraction of longitudinal momentum will be smaller by
a factor $\varepsilon_N/E_F$:
\begin{equation}
\int_0^1\!\!
\!\!F^{N}_{2}\!\left(\frac{E_F}{\varepsilon_N}x_N\!\!\right)\!dx_N\!
=\frac{\varepsilon_N}{E_F}\!\!\int_0^{\frac{E_F}{\varepsilon_N}}\!\!
F^{N}_{2}\!\left(x\right)\!dx\!\cong\!
\frac{\varepsilon_N}{E_F}\!\!\int_0^{1}\!\!\!
\!F^{N}_{2}\!\left(x\right)\!dx \label{MSR1}
\end{equation}
Here in  the integral we neglect the small contributions from
$x>1$ region originated from NN correlations.
\begin{figure}
\includegraphics[height=8cm,width=8.5cm]{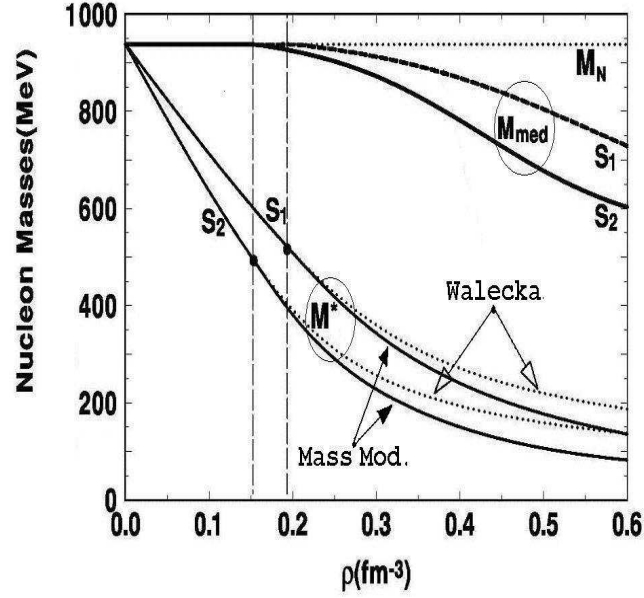}
\caption{Constant nucleon mass $M_N$ (Walecka) and density
dependent mass $M_{med}$ from our "Mass Mod." model. Also
effective mass $M^*$ in these approaches. Both models are
calculated for $S_1$ and $S_2$ parametrization.} \label{fig:mass}
\end{figure}

Now, with the help of Eq.(\ref{RMF2}) and Eq.(\ref{MSR1}), the
nuclear MSR is satisfied:
\begin{eqnarray}
&\frac{1}{A}\int\!F^{A}_{2}(x_A)dx_A = \int\!dy y \rho ^{A}(y)
\int\!
F^{N}_{2}\left(\frac{M_N}{M_{mod}}x_N\right)dx_N\cong \nonumber \\
&\frac{E_F}{\varepsilon_N}\frac{\varepsilon_N}{E_F}\int\!
F^{N}_{2}(x)dx=\int\! F^{N}_{2}(x)dx.
\end{eqnarray}

 This means
that quarks in the nucleus carry the same fraction of longitudinal
momentum as in bare nucleons.

On the other hand the integral (\ref{MSR1}) corresponds (\ref{eq})
to the total sum of the quark longitudinal momenta
$p_q^+=p_q^0+p_q^3$ inside a nucleon, which is proportional to a
total nucleon rest energy or the nucleon mass. Consequently, the
nucleon mass $M_N$ will be changed for $\varrho\geq\varrho_0$ to
the mass $M_{med}$ by the gradually decreasing factor
$\varepsilon_N/E_F$:

\begin{equation}
M_{med}\!=\!\frac{\varepsilon_N}{E_F}M_N\!=\!M_N\!/
\left(1\!+\!\frac{\varrho\frac{d}{d\varrho}\left({\varepsilon_N}\right)}{
\varepsilon_{N} }\right)\!\simeq\!M_N\!\left(
1\!-\!\frac{p}{\varrho\varepsilon_{N}} \right) \label{masseff}
\end{equation}
which decreases as the pressure increases. This explicit mass
dependence from energy $\varepsilon_N$ and energy derivative
(\ref{masseff}) is plugged into the following standard Walecka RMF
equations\cite{wa} for energy per nucleon $\varepsilon_N$ and
effective mass $M^*$:
\begin{figure}
\vspace{-15mm}
\includegraphics[height=8.5cm,width=12.cm]{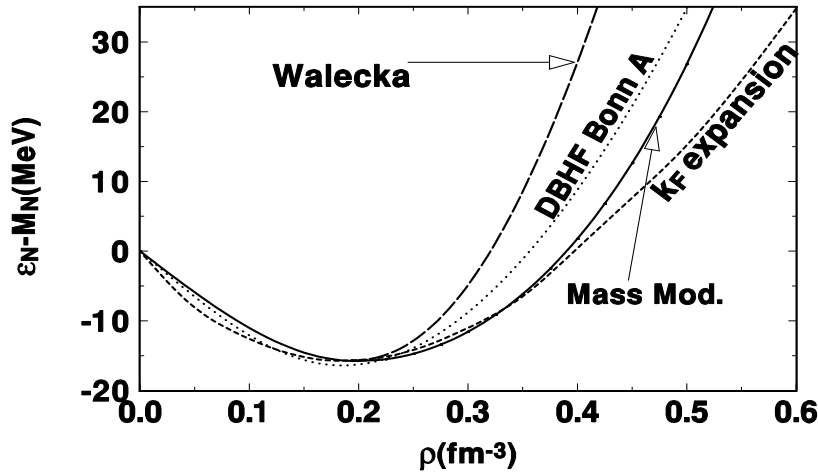}
\caption{The nucleon energy $\varepsilon_N-M_N$ as a function of
NM density for two RMF models; scalar-vector Walecka (dot lines)
and our Modified Mass approach (solid). Both RMF models are
calculated for two parameterizations: $S_1$ version\cite{wa}
$(\rho_0=.19fm^{-3})$ and version\cite{ser} $S_2$
$(\rho_0=.16fm^{-3}$). Results for full DBHF\cite{bonn} (dotted
marked line) calculation using Bonn A NN interaction are displayed
for comparison, also nucleon energy in ZM model\cite{zm} is in the
plot (dotted marked line).} \label{fig:enernew}
\end{figure}

\section{Conclusion}

 Here, our model applied to the
linear Walecka model for nuclear and neutron matter make the EoS
softer, close to semi-empirical analysis\cite{pawel} and close to
DBHF calculation with a realistic Bonn A potential\cite{bonn}.
Other features of the Walecka model,
 including a good value of the spin-orbit force remain in our model
 unchanged.
Our results suggest  corrections above the saturation density to
any RMF model, with constant nucleon mass and unmodified parton
SF. The stiffness of EoS recently discussed\cite{NSTARS} in
application to compact and neutron stars is important when
studying star properties (mass-radius constraint). Partial support
of the Ministry of Science and Higher Education under  the
Research Project No. N N202046237 for is acknowledged.

\end{document}